\documentclass[sigconf,authorversion]{acmart}

\AtBeginDocument{%
  \providecommand\BibTeX{{%
    \normalfont B\kern-0.5em{\scshape i\kern-0.25em b}\kern-0.8em\TeX}}}

\setcopyright{acmcopyright}
\copyrightyear{2019}
\acmYear{2019}
\acmDOI{10.1145/3331184.3331366}

\acmConference[SIGIR '19] {42nd International ACM SIGIR Conference on Research and Development in Information Retrieval}{July 21--25, 2019}{Paris, France}
\acmBooktitle{42nd Int'l ACM SIGIR Conference on Research and Development in Information Retrieval (SIGIR '19), July 21--25, 2019, Paris, France}
\acmPrice{15.00}
\acmISBN{978-1-4503-6172-9/19/07}

\usepackage{booktabs} % For formal tables
\usepackage{csquotes}
\usepackage{paralist}

\makeatletter
\g@addto@macro{\UrlBreaks}{\do\-}
\makeatother 

\begin{document}

\title{Retrieving Multi-Entity Associations: An~Evaluation~of~Combination~Modes~for~Word~Embeddings}

	\author{Gloria Feher}
	\affiliation{%
	  \institution{Heidelberg University}
	  \streetaddress{Im Neuenheimer Feld 205}
	  \city{Heidelberg}
	  \country{Germany}
	  \postcode{69120}
	}
	\email{gloria.e.feher@gmail.com}
	
	\author{Andreas Spitz}
	\affiliation{%
		\institution{Heidelberg University}
		\streetaddress{Im Neuenheimer Feld 205}
		\city{Heidelberg}
		\country{Germany}
		\postcode{69120}
	}
	\email{spitz@informatik.uni-heidelberg.de}
	
	\author{Michael Gertz}
	\affiliation{%
		\institution{Heidelberg University}
		\streetaddress{Im Neuenheimer Feld 205}
		\city{Heidelberg}
		\country{Germany}
		\postcode{69120}
	}
	\email{gertz@informatik.uni-heidelberg.de}

%\renewcommand{\shortauthors}{Feher et al.}

\begin{abstract}
Word embeddings have gained significant attention as learnable representations of semantic relations between words, and have been shown to improve upon the results of traditional word representations. 
However, little effort has been devoted to using embeddings for the retrieval of entity associations beyond pairwise relations. 
In this paper, we use popular embedding methods to train vector representations of an entity-annotated news corpus, and evaluate their performance for the task of predicting entity participation in news events versus a traditional word cooccurrence network as a baseline. 
To support queries for events with multiple participating entities, we test a number of combination modes for the embedding vectors. 
While we find that even the best combination modes for word embeddings do not quite reach the performance of the full cooccurrence network, especially for rare entities, we observe that different embedding methods model different types of relations, thereby indicating the potential for ensemble methods. 
\end{abstract}

\begin{CCSXML}
		<ccs2012>
		<concept>
		<concept_id>10002951.10003317.10003338</concept_id>
		<concept_desc>Information systems~Retrieval models and ranking</concept_desc>
		<concept_significance>500</concept_significance>
		</concept>
		<concept>
		<concept_id>10002951.10003317.10003359.10003362</concept_id>
		<concept_desc>Information systems~Retrieval effectiveness</concept_desc>
		<concept_significance>500</concept_significance>
		</concept>
		</ccs2012>
\end{CCSXML}
	
\ccsdesc[500]{Information systems~Retrieval models and ranking}
\ccsdesc[500]{Information systems~Retrieval effectiveness}

\keywords{word embeddings; embedding vector combination; implicit network; entity network}


\maketitle

%%%%%%%%%%%%%%%%%%
% BEGIN: CONTENT
%%%%%%%%%%%%%%%%%%

\section{Introduction}

Word embeddings are learned dense vector representations of words, which encode information on word context. They have established themselves as a popular way to encode unstructured text due to their numerous useful properties, such as the clustering of semantically or syntactically related words in the vector space, or the support of arithmetic operations on word vectors to \enquote{calculate} word analogies. These characteristics lend themselves to tasks in natural language processing and information retrieval, where embeddings can be used to alleviate vocabulary mismatch~\cite{ganguly2015word}, for example. 
Similarly, sensitivity classification~\cite{conf/ecir/McDonaldMO17} and large-scale text classification~\cite{journals/corr/BalikasA16a} have been improved by using embeddings, although only in combination with other task-relevant features. 
It is thus important to investigate when, how, and why word embeddings perform so well, and when they do not. 
In particular, the issue has been raised that word embeddings provide meaningless similarities between otherwise unrelated words if the entire vector space is considered~\cite{conf/ecir/KarlgrenHS08}, and it has been demonstrated that local embeddings outperform global embeddings for query expansion~\cite{conf/acl/0001MC16}. 
These considerations imply that the potential neighbourhood in the word embedding space is too large and needs to be restricted in order to mimic only meaningful relations, which raises some questions such as: 
\begin{inparaenum}[(1)]
  \item Do word embeddings universally capture the relevant associations encoded in language better than other methods?
  \item How is the neighbourhood of embeddings best used to solve a task pertaining to non-trivial word associations?
\end{inparaenum}	

To investigate these questions, we consider 
the task of event completion, where one held-out entity is predicted from the remaining entities that participate in an event. An entity is said to participate in an event, if it is named in its description. Thus, predicting one entity from other participating entities is a suitable problem to evaluate relevant associations between words as captured by word embeddings, as well as different combination modes of word vectors to exploit the neighbourhood relationships. Since the task relies on the cooccurrence of entities in a common context, it lends itself to the use of embeddings. However, the fact that an entity may occur in different contexts provides a challenge for learned word vectors (for example, Brazil held the Summer Olympics in 2016, and in the same year impeached its president Dilma Rousseff for breaking fiscal laws). 
Benchmarks and training data for such an event completion task are provided by Spitz and Gertz~\cite{DBLP:conf/www/SpitzG18}, who used it to evaluate ranking in entity cooccurrence networks. 
Research by Schnabel et al.\ suggests that there is no universally best embedding method since the performance of embeddings varies by task~\cite{conf/emnlp/SchnabelLMJ15}. We therefore evaluate word2vec-CBOW, word2vec-skip-gram, and GloVe to determine which method is best suited for identifying the participation of entities in events, and discuss the benefits and drawbacks of each tested method.

\noindent\textbf{Contributions.} 
We make two primary contributions.
\begin{inparaenum}[(1)]
	\item We compare six modes of combining embedding vectors for multi-entity queries.
	\item Based on a comparison to entity ranking in cooccurrence networks, we discuss the influence of an entity's frequency in the corpus on the performance of its resulting embedding. 
\end{inparaenum}

\section{Related Work}
Word embeddings are distributional vector representations of words that have been researched since the 1990s~\cite{elman:distributed}, leading to methods such as Latent Dirichlet Allocation~\cite{Blei03ldaModel} and neural language models~\cite{bengio2003neural}. These provide the groundwork for current word embedding models, which usually employ shallow neural networks. Recently, the context-based embeddings ELMo~\cite{DBLP:conf/naacl/PetersNIGCLZ18} and BERT~\cite{DBLP:journals/corr/abs-1810-04805} have been introduced, which are generated by data and training intensive deep architectures. These word representations vary with the input sentence(s), and are thus unsuitable for an isolated entity retrieval task, so we rely on more traditional embeddings, which yield only one fixed vector per word.

\noindent\textbf{Word2vec} 
is one of the most widely used models and comes in two variants~\cite{mikolov2013efficient,mikolov2013distributed}. Continuous bag-of-words (CBOW) averages the word vectors of context words and uses them to classify the focal word. Inversely, continuous skip-gram predicts the context words from the focal word. Thus, word vectors are learned from local context windows. Typically, word2vec is modified to include negative sampling~\cite{mikolov2013distributed}, which distinguishes between the real context words and a sample that is drawn from a noise distribution.

\noindent\textbf{GloVe} 
was proposed
as global word representation vectors that improve upon the approach of word2vec~\cite{DBLP:conf/emnlp/PenningtonSM14}. To this end, they make use of the global corpus statistics by training a log-bilinear regression model on the word cooccurrence matrix. By employing a more general weighting function, they also gain more control over how strongly frequent words influence the model.

\noindent\textbf{Implicit networks}, in contrast, 
were introduced
to model latent relationships between entities 
as well as the remaining terms~\cite{DBLP:conf/sigir/SpitzG16}. The latent relationships are captured by constructing an entity and term coocurrence graph that is weighted by cross-sentence distances within the documents, thus capturing all word cooccurrences in the text. While the model has a variety of applications like event extraction or entity-centric topic extraction~\cite{DBLP:conf/ecir/SpitzG18}, it comes with event completion ground-truth data for news articles~\cite{DBLP:conf/www/SpitzG18}, so we use it as a baseline. With the prevalence of machine learning applications, it is also of interest to evaluate how well the implicit network performs in comparison to learned representations.

\noindent\textbf{Combining word vectors}
as a technique is typically employed for creating document embeddings from word embeddings~\cite{journals/corr/BalikasA16a,DBLP:conf/www/NalisnickMCC16} or for modeling multi-word entities in knowledge bases~\cite{DBLP:conf/nips/SocherCMN13}. Several combination modes exist in the literature, mainly component-wise minimum and maximum, 
as well as averaging~\cite{journals/corr/BalikasA16a,DBLP:conf/nips/SocherCMN13,DBLP:conf/nips/SocherHPNM11}, which are often combined by concatenation into a single vector. Another approach is summing over the similarities between multiple query vectors and document vectors~\cite{DBLP:conf/www/NalisnickMCC16}. Mitchell and Lapata compare several composition functions for handcrafted cooccur\-rence-based word vectors~\cite{DBLP:journals/cogsci/MitchellL10}. Iacobacci et al. compare word embedding combinations for the task of word sense disambiguation~\cite{DBLP:conf/acl/IacobacciPN16}. However, to the best of our knowledge, no previous publication has compared different combination modes of learned word vectors obtained from several embedding methods on an entity-centric task.

\section{Experimental Setup} 

We briefly describe the task and training data, as well as the necessary steps for tuning the used embeddings. 

\noindent\textbf{Event completion task.}
To formalize the task, we assume that an event is defined by its $k$ participating entities of type location, person, or organization. For each event, we generate $k$ queries by holding-out one entity and using the remaining $k-1$ entities as query input. The task is then to predict the held-out entity, i.e., given the query entities and the type of the target entity, each model should predict the held-out entity.
In practice, we treat this as a ranking task in which we rank nearest neighbours by cosine distance (for embeddings) or adjacent nodes by edge weights (for the implicit network). For each query, we generate a ranking of target entities to calculate $precision@1$ and $recall$.
For example, the event of the third presidential debate of the 2016 U.S.\ election between Hillary Clinton (\textsc{HC}) and Donald Trump (DT) at the University of Nevada, Las Vegas (UNLV), would yield the three queries 
$\langle \{\textrm{\textsc{hc}},\textrm{\textsc{dt}}\} \to \{\textrm{\textsc{unlv}}\} \rangle$, 
$\langle \{\textrm{\textsc{hc}},\textrm{\textsc{unlv}}\} \to \{\textrm{\textsc{dt}}\} \rangle$, and 
$\langle \{\textrm{\textsc{dt}},\textrm{\textsc{unlv}}\} \to \{\textrm{\textsc{hc}}\} \rangle$.

\noindent\textbf{Data.}
Since event completion is most relevant on news data, we train all models on a corpus of 127,485 English news articles from June 2016 to November 2016~\cite{DBLP:conf/www/SpitzG18},  
and follow the preprocessing steps described there to construct the implicit network. To train the embeddings, we replace named entity mentions in the text with Wikidata IDs. 
As ground truth, we use data from the same source~\cite{DBLP:conf/www/SpitzG18}, which was accumulated by crawling the Wikipedia Current Events portal  
for events that are described in articles within the corpus, and removing events that contain less than two entities. For our evaluation, we also exclude all queries in which the target entity is missing in at least one of our evaluated models (due to different window sizes and retention threshold values, not all models contain all entities). The final ground truth contains 263 queries. 

\noindent\textbf{Parameter tuning.} 
We conduct preliminary hyperparameter optimization for all embedding models and select the best settings per method according to the achieved $precision@1$ scores. All embeddings are trained for 100 epochs, and we consider embeddings of dimension 50, 100, and 200. Where not otherwise specified, we choose hyperparameters according to the recommended default values. Due to the non-deterministic nature of word2vec and GloVe, we train each embedding ten times and average the results. Training is time intensive and takes up to 18h per model on a dual-core machine for skip-gram models.

\noindent\textbf{Multi-entity neighbourhood.}  
To use embeddings to predict the target entity, we propose a new mode for combining the word vectors of individual query entities and test five further modes from the literature. Let $Q$ 
denote the set of query entities.
Let $x \in X$ be one out of all possible target entities, with $\theta_x$ denoting its word vector, and $\Theta^Q = [\theta_{q1}, \theta_{q2}, \cdots]$ the matrix of horizontally stacked word vectors of query entities. Then $t_\textsc{mode} \in X$ is the predicted target entity according to the respective ranking functions.
\begin{align}
t_\textsc{minmax} &= \underset{x \in X}{\mathrm{argmin}} \quad \underset{q \in Q}{\mathrm{argmax}} \quad cosdist ( \theta_q,\theta_x ) \label{eq:minmax}\\
t_\textsc{sum} &= \underset{x \in X}{\mathrm{argmin}} \sum_{q \in Q} cosdist ( \theta_q,\theta_x ) \label{eq:sum}\\
t_\textsc{avg} &= \underset{x \in X}{\mathrm{argmin}} \quad cosdist \Bigl( \frac{1}{|Q|} \sum_{q \in Q} \theta_q, \theta_x \Bigr) \label{eq:avg}\\
t_\textsc{cwmin} &= \underset{x \in X}{\mathrm{argmin}} \;\;\, cosdist \Bigl([\mathrm{min}(\Theta^Q_1), \cdots, \mathrm{min}(\Theta^Q_{|Q|})]^T, \theta_x  \Bigr)\label{eq:cwmin}
\end{align}
\begin{align}
t_\textsc{cwmult} &= \underset{x \in X}{\mathrm{argmin}} \;\;\, cosdist \Bigl(\theta_{q_1} \odot \cdots \odot \theta_{q_{|Q|}}, \quad \theta_x  \Bigr), q_i \in Q\label{eq:cwmult}
\end{align}
Thus, MINMAX (Eq. \ref{eq:minmax}) finds the word vector with the minimal maximal distance to the query vectors, SUM (Eq. \ref{eq:sum}) finds the word vector with the minimal sum of cosine distances to all query vectors~\cite{DBLP:conf/www/NalisnickMCC16}, and AVG (Eq. \ref{eq:avg}) first averages the query vectors and then finds the nearest neighbour~\cite{journals/corr/BalikasA16a}. We further denote finding the word vector with the smallest cosine distance to the component-wise minimum of all query vectors with CWMIN (Eq. \ref{eq:cwmin})~\cite{journals/corr/BalikasA16a}, the component-wise maximum with CWMAX (analogous to Eq. \ref{eq:cwmin}) \cite{journals/corr/BalikasA16a} and the component-wise multiplication with CWMULT (Eq. \ref{eq:cwmult})~\cite{DBLP:journals/cogsci/MitchellL10}.

Note that these combination modes are especially suitable for applications that do not depend on word order or syntactical information, since all of the evaluated modes are symmetrical. The event completion task is agnostic to word order as it operates on sets of entities. This is also true for implicit networks, which establish entity relations based on (symmetric) sentence-distance.

\section{Evaluation}

\begin{table}[b]
	
	\vspace*{-5pt}

	\caption{Average $prc@1$ for the event completion task for all modes and all embedding models of size 200. The implicit network baseline achieves an average $prc@1$ of $0.330$.}
	
	\vspace*{-5pt}
	
	\begin{tabular}{lccc}
		\toprule
		& skip-gram & CBOW & GloVe \\ %& LOAD \\
		\midrule
%		& & & & 0.330\\
		SUM & 0.257 & 0.234 & 0.252    \\% & \\
		AVG & 0.140 & 0.116 & 0.101    \\% & \\
		MINMAX & 0.189 & 0.186 & 0.168 \\% & \\
		CWMAX & 0.140 & 0.102 &  0.085 \\% & \\
		CWMIN & 0.130 & 0.095 & 0.095  \\% & \\
		CWMULT & 0.085 & 0.066 & 0.056 \\% & \\
		\bottomrule
	\end{tabular}
	\label{tab:rec}%
\end{table}

\begin{figure}[t]
	
	\centering
	\includegraphics[width=\linewidth, height=70pt]{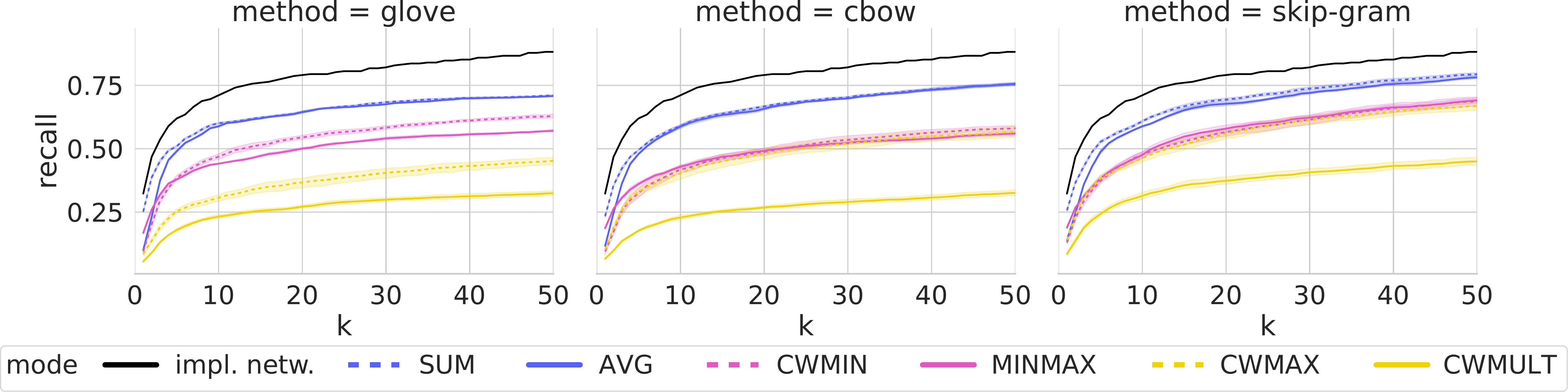}
	
	\vspace*{-5pt}
	
	\caption{$Recall@k$ for GloVe, CBOW, and skip-gram embeddings with the combination modes SUM, MINMAX, AVG, CWMIN, CWMAX and CWMULT. The implicit network is included as a baseline.}
	\label{fig:rec}
\end{figure}

In order to compare the word vector combination modes, we first focus our evaluation on their performance in the event completion task. Figure~\ref{fig:rec} shows the $recall@k$ for all combination modes and embedding methods. We observe that the combination modes perform similarly across the different embeddings, where both SUM and AVG perform comparably well in terms of recall, albeit worse than the implicit network baseline. MINMAX, CWMIN, CWMAX and CWMULT perform considerably worse and do not reach a recall above $0.6$ at rank 10, unlike the other combination modes. In terms of $precision@1$, SUM significantly outperforms the other modes across all embedding methods, as is further highlighted in Table~\ref{tab:rec}. We thus focus on SUM as a combination mode in the following.

\begin{figure}[t]
	\vspace*{-10pt}
			\centering
			\includegraphics[width=0.8\linewidth]{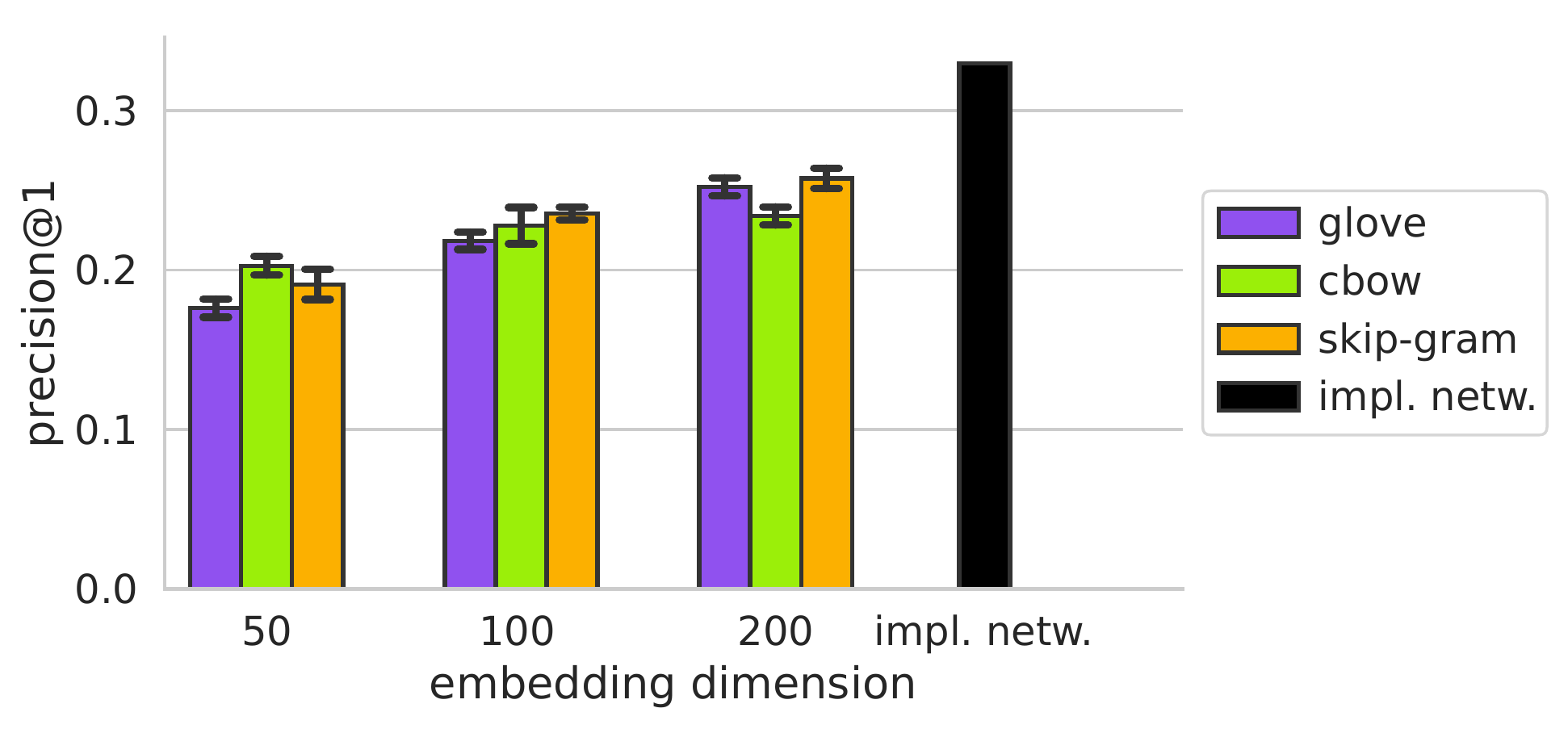}
			\vspace*{-5pt}
			\caption{Average $prc@1$ for CBOW, skip-gram, and GloVe models with dimension 50, 100, 200 (using mode SUM).}%
			\label{fig:prec}
	\vspace*{-10pt}
\end{figure}

In Figure~\ref{fig:prec}, we show the $precision@1$ for the SUM mode and the embedding models that perform best out of all tested hyperparameter settings, to compare embedding spaces with differing dimensionalities.
For CBOW, we obtain the best performance for a down-sampling threshold of $10^{-5}$, $15$ negative samples, a context window size of $21$, and a minimum of 3 word occurrences in the training data. Similarly, skip-gram performs best for a down-sampling threshold of $10^{-5}$, and a window size of $21$. Like the word2vec models, GloVe yields the best results for a window size of 21, as well as setting $\alpha = 0.6$ and $x_{max} = 25$ in the weighting function of the least squares objective~\cite{DBLP:conf/emnlp/PenningtonSM14}.

Overall, we observe two general trends. First, increasing the embedding dimension almost universally improves the performance. Second, using a larger window size of $21$ yields improvements across all methods (compared to the recommended values of $15$ for GloVe, $10$ for skip-gram, and $5$ for CBOW). This indicates the importance of considering larger contexts to capture relevant relations between entities that are typically farther apart in a text, and is further corroborated by the implicit network's increased performance as it considers all relations within a document.

\begin{figure}[t]
	\centering
	
	\includegraphics[width=0.61\linewidth]{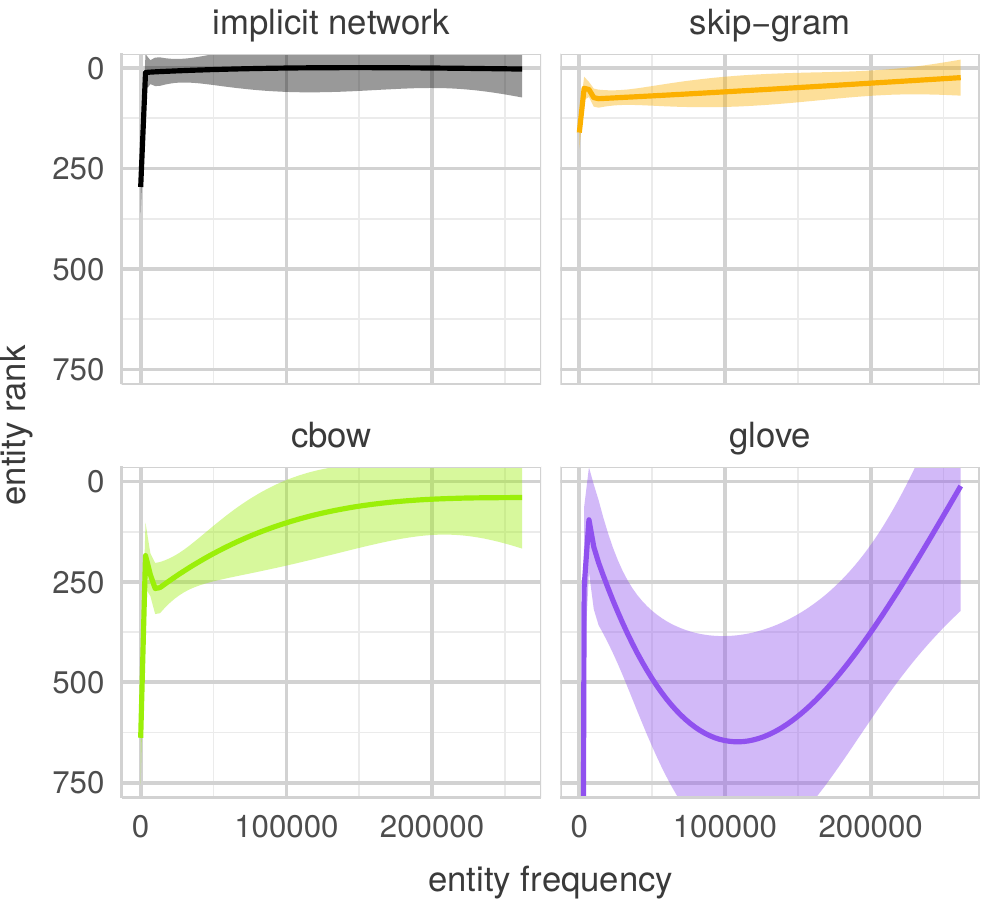}
	\vspace*{-6pt}
	\caption{Obtained ranks of target entities for all methods vs.\ the frequency of target entities in the training corpus. Shaded areas denote $0.95$ confidence intervals.}
	\label{fig:freq}
	
\end{figure}

Another interesting aspect to consider is the frequency of target entities in the training data, since it should be more difficult to train models for rare entities. As the results in Figure~\ref{fig:freq} show, the implicit network unsurprisingly performs best for rare entities as it is constructed to retain the full, uncompressed cooccurrence information. However, while skip-gram and CBOW both benefit from higher entity frequencies, skip-gram initially performs best (even better than the implicit network) for extremely rare entities (frequency $\leq 10$). GloVe produces the most curious results and performs worst for only some entities with very low frequencies, but also for those with mid-range frequencies, despite an overall performance that is close to skip-gram, as shown in Figure~\ref{fig:prec}.

\begin{figure}[t]
	\centering
	
	\vspace*{-5pt}		
	
	\includegraphics[width=\linewidth]{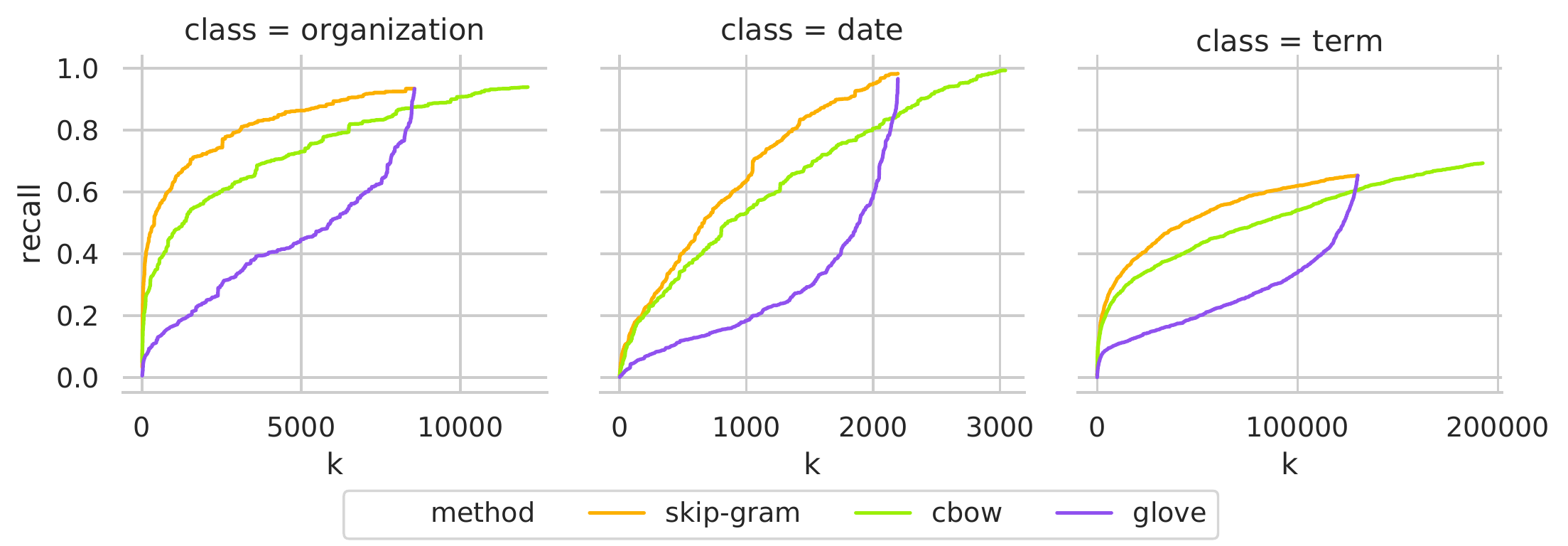}
	\vspace*{-17pt}
	\caption{Average recall curves of predictions by all embedding methods (using SUM) for the top 100 neighbours of 25 randomly selected entities in the implicit network.}
	\label{fig:rel_rec}
	
	\vspace*{-8pt}
	
\end{figure}

Due to the similar overall performance of GloVe and skip-gram in contrast to the apparent difference in performance depending on entity frequencies, we also investigate whether the models capture similar relations between entities in comparison to the implicit network. We randomly select 25 entities of each type
and use the 100 most closely related entities in the implicit network as a pseudo ground-truth against which we rank the predictions that are generated by the embeddings. 
In Figure~\ref{fig:rel_rec}, we show the recall curves for the three embedding models. Note that the recall does not always reach a value of $1$ since some rare entities are not contained in all embeddings due to the window size constraints that are more strict than in the implicit network. 
Here, we find that the word2vec models appear to rank entities in a similar manner to the implicit network, whereas GloVe ranks them differently and initially has lower recall. However, since GloVe performs similarly well on the event completion task, this indicates that GloVe embeddings model a different sort of relation between the involved entities of events that nevertheless results in a similar overall performance.

\section{Conclusion and Ongoing Work}

For our investigation into the combination modes of word vectors, we conclude that summing over the cosine distances between the query vectors and prospective target vectors yields the highest precision and recall in the event completion task. While we focused on entities in our evaluation due to the availability of suitable training, query, and ground truth data, the experimental setup is generic. Therefore, we expect the results to be similar for multi-word associations in general once respective data sets become available. 

Due to their compact representation and computational efficiency, learned word vectors are appealing for many applications. 
However, for the semantic-relatedness task of retrieving multi-entity associations, our experiments show that the representations obtained from word2vec and GloVe do not yet reach the performance of a full cooccurrence network representation, even though they are close. This, of course, raises the question why the learned embeddings do not attain the same performance level as the heuristic representation, and how this can be addressed in the future.

One potential explanation for the performance gap is the sparseness of entity mentions in comparison to general terms. Where the implicit network contains edges that represent cooccurrences over distances of multiple sentences, the embeddings are limited by a stricter window size, which conforms with their increased performance for increasing window size. However, increasing the window size does not scale arbitrarily due to computational restrictions on the runtime and the introduction of noise. While entities are bound to share some relation across sentence boundaries, this is less likely to be the case for arbitrary terms. Here, combination methods for training embeddings stand to improve the overall performance.

Our final observation concerns the fact that GloVe captures inherently different entity associations when compared to both word2vec and the implicit network. While implicit networks may generally capture more of the useful entity relations, this difference indicates that they apparently miss some of the associations that are emphasized in the global embedding process. A further analysis of these associations and how they occur would be of interest. Combining the network-based model with multiple learned features in an ensemble approach may benefit the overall performance, and requires a more in-depth investigation of the differences between the underlying relationships of entities in the different methods.

%%%%%%%%%%%%%%%%%%
% END: CONTENT
%%%%%%%%%%%%%%%%%%

\bibliographystyle{ACM-Reference-Format}
\bibliography{bibliography-included.bib}

\end{document}